\documentclass[twoside,a4paper]{amsart}
\usepackage[T1,T2A]{fontenc}
\usepackage[koi8-r]{inputenc}
\usepackage[english,russian]{babel}
\newcommand{\cprime}{'}
\providecommand{\text}[1]{\mbox{#1}}

\newcommand{\AMSMSC}[2]{\subjclass[2010]{Primary #1; Secondary #2.}}
\providecommand{\subjclass}[2][]{}

\usepackage[pdfauthor={Vladimir V. Kisil},%
    pdftitle={Classical/Quantum=Commutative/Noncommutative?},%
    pdfsubject={mathematics, physics},%
    backref=page,%
  pdfkeywords={symmetries, quantum mechanics, hypercomplex numbers},
breaklinks=true,%
colorlinks=true,%
bookmarks=true,%
backref=page,%
pagebackref=true
]{hyperref}
\usepackage[backrefs,msc-links
,nobysame
]{myamsrefs}

   \PII{\relax}
   \copyrightinfo{}{\relax}

\providecommand{\Space}[3][]{\ensuremath{\mathbb{#2}^{#3}_{#1}{}}}
  \providecommand{\FSpace}[3][]{\ensuremath{\ifx#2l \ell_{#3}^{#1}{}\else
  #2_{#3}^{#1}{}\fi}} 
\providecommand{\myh}{h}
\providecommand{\myhbar}{\hslash}
\providecommand{\rmi}{\mathrm{i}}

\providecommand{\rmh}{\mathrm{j}}
\providecommand{\rmp}{\varepsilon}
\providecommand{\comment}[1]{}
\providecommand{\uir}[3][0]{\ifcase #1{\rho^{#2}_{#3}}%
\or {\breve{\rho}^{#2}_{#3}}%
\or {\tilde{\rho}^{#2}_{#3}}\fi}
\providecommand{\algebra}[1]{\ensuremath{\mathfrak{#1}}}
  \providecommand{\Zbl}[1]{Zbl\href{http://www.emis.de:80/cgi-bin/zmen/ZMATH/en/zmathf.html?first=1&maxdocs=3&type=html&an=#1&format=complete}{#1}}

\providecommand{\myeprint}[2]{E-print: \href{#1}{\texttt{#2}}}
\providecommand{\doi}[1]{doi: \href{http://dx.doi.org/#1}{#1}}

\begin{document}
  \selectlanguage{english} 
\title[Classical/Quantum=Commutative/Noncommutative?] %
{Classical\(/\)Quantum\(=\)Commutative\(/\)Noncommutative?}

\author[Vladimir V. Kisil]%
{\href{http://www.maths.leeds.ac.uk/~kisilv/}{Vladimir V. Kisil}}
\thanks{School of Mathematics,
University of Leeds,
Leeds LS2\,9JT,
UK.}
\thanks{email: \href{mailto:kisilv@maths.leeds.ac.uk}{\texttt{kisilv@maths.leeds.ac.uk}}.}
\thanks{Web: \url{http://www.maths.leeds.ac.uk/~kisilv/}.}

\thanks{On  leave from Odessa University.}

\address{%
School of Mathematics\\
University of Leeds\\
Leeds LS2\,9JT\\
UK
}

\email{\href{mailto:kisilv@maths.leeds.ac.uk}{kisilv@maths.leeds.ac.uk}}

\urladdr{\href{http://www.maths.leeds.ac.uk/~kisilv/}%
{http://www.maths.leeds.ac.uk/\~{}kisilv/}}

\date{\today}

\begin{abstract}
  In 1926, Dirac stated  that quantum mechanics can be
  obtained from classical theory through a change in the only rule. In his
  view, classical mechanics is formulated through commutative
  quantities (c-numbers) while quantum mechanics requires
  noncommutative one (q-numbers). The rest of theory can be
  unchanged. In this paper we critically review Dirac's proposition.

  We provide a natural formulation of classical mechanics through
  noncommutative quantities with a non-zero Planck constant. This is
  done with the help of the nilpotent unit \(\rmp\) such that
  \(\rmp^2=0\). Thus, the crucial r\^ole in quantum theory shall be
  attributed to the usage of complex numbers.
\end{abstract}
\AMSMSC{81P05}{22E27}
\maketitle

\par
\hfill\parbox{0.6\textwidth}{\footnotesize\ldots it was on a Sunday
  that the idea first occurred to me that \(ab- ba\) might correspond to a
  Poisson bracket. 
 \par \hfil P.A.M. Dirac,
}\par\medskip



There is a recent revival of interest in foundations of quantum
mechanics, 
which is essentially motivated by engineering challenges at the
nano-scale. There are strong indications that we need to revise the
development of the quantum theory from its early days.

In 1926, Dirac discussed the idea that quantum mechanics can be obtained
from classical one through a change in the only rule, cf.~\cite{Dirac26a}:
\begin{quote}
  \ldots there is one basic assumption of the classical theory which
  is false, and that if this assumption were removed and replaced by
  something more general, the whole of atomic theory would follow
  quite naturally. Until quite recently, however, one has had no idea
  of what this assumption could be.
\end{quote}

In Dirac's view, such a condition is provided by the Heisenberg commutation
relation of coordinate and momentum variables~\cite{Dirac26a}*{(1)}:
\begin{equation}
  \label{eq:heisenberg-comm-basic}
  q_r p_r-p_r q_r=\rmi \myh.
\end{equation}
Algebraically, this identity declares noncommutativity of \(q_r\) and
\(p_r\). Thus, Dirac stated~\cite{Dirac26a} that classical mechanics is formulated
through commutative quantities (``c-numbers'' in his terms) while
quantum mechanics requires noncommutative quantities (``q-numbers''). The
rest of theory may be unchanged if it does not contradict to the above
algebraic rules. This was explicitly re-affirmed at the
first sentence of the subsequent paper~\cite{Dirac26b}:
\begin{quote}
  The new mechanics of the atom introduced by Heisenberg may be based
  on the assumption that the variables that describe a dynamical
  system do not obey the commutative law of multiplication, but
  satisfy instead certain quantum conditions.
\end{quote}
The same point of view is expressed in his later works \citelist{\cite{DiracDirections}*{p.~6}
\cite{DiracPrinciplesQM}*{p.~26}}.

Dirac's approach was largely approved, especially by researchers
on the mathematical side of the board. Moreover, the vague version
``quantum is something noncommutative'' of the original statement
 was lightly reverted to ``everything
noncommutative is quantum''. For example, there is a fashion to label
any noncommutative algebra as a ``quantum space''~\cite{Cuntz01a}. 

Let us carefully review Dirac's  idea about noncommutativity as the
principal source of quantum theory.

\section{``Algebra'' of Observables}
\label{sec:algebra-observables}

Dropping the commutativity hypothesis on observables, Dirac
made~\cite{Dirac26a} the following (apparently flexible) assumption:
\begin{quote}
  All one knows about q-numbers is that if \(z_1\) and \(z_2\) are two
  q-numbers, or one q-number and one c-number, there exist the numbers
  \(z_1 + z_2\), \(z_1 z_2\), \(z_2 z_1\), which will in general be
  q-numbers but may be c-numbers.
\end{quote}
Mathematically, this (together with some natural identities) means
that observables form an algebraic structure known as a \emph{ring}.
Furthermore, the linear \emph{superposition principle} imposes a liner
structure upon observables, thus their set becomes an \emph{algebra}.
Some mathematically-oriented texts,
e.g.~\cite{FaddeevYakubovskii09}*{\S~1.2}, directly speak about an
``algebra of observables'' which is not far from the above
quote~\cite{Dirac26a}. It is also deducible from two connected
statements in Dirac's canonical textbook:
\begin{enumerate}
\item ``the linear operators corresponds to the dynamical variables at
  that time''~\cite{DiracPrinciplesQM}*{\S~7, p.~26}.
\item ``Linear operators can be added
  together''~\cite{DiracPrinciplesQM}*{\S~7, p.~23}. 
\end{enumerate}

However, the assumption that any two observables may be added cannot
fit into a physical theory. To admit addition, observables need to
have the same dimensionality. In the simplest example of the
observables of coordinate \(q\) and momentum \(p\), which units shall
be assigned to the expression \(q+p\)? Meters or
\(\frac{\text{kilos}\times\text{meters}}{\text{seconds}}\)? If we get the value \(5\)
for \(p+q\) in the metric units, what is the result in the imperial
ones? 
Since these questions cannot be answered, the above
Dirac's assumption is not a part of any physical theory.

Another common definition suffering from the same problem is used in
many excellent books written by distinguished mathematicians, see for
example \citelist{\cite{Mackey63}*{\S~2-2} \cite{Folland89}*{\S~1.1}}.
It declares that quantum observables are projection-valued Borel
measures on the \emph{dimensionless} real line. Such a definition
permit an instant construction (through the functional
calculus) of new observables, including
algebraically formed~\cite{Mackey63}*{\S~2-2, p.~63}:
\begin{quote}
  Because of Axiom III, expressions such as \(A^2\), \(A^3+A\),
  \(1-A\), and \(e^A\) all make sense whenever \(A\) is an
  observable. 
\end{quote}
However, if \(A\) has a dimension (is not a scalar) then the
expression \(A^3+A\) cannot be assigned a dimension in a consistent
manner.

Of course, physical defects of the above (otherwise perfect)
mathematical constructions do not prevent physicists from making
correct calculations
, which are in a good
agreement with experiments. We are not going to analyse methods which
allow researchers to escape the indicated dangers. Instead, 
it will be more beneficial to outline alternative mathematical
foundations of quantum theory, which do not have those shortcomings.

\section{Non-Essential  Noncommutativity}
\label{sec:essent-noncomm}

While we can add two observables if they have the same dimension only,
physics allows us to multiply any observables freely. Of course, the
dimensionality of a product is the product of dimensionalities, thus
the commutator \([A,B]=AB-BA\) is well defined for any two observables
\(A\) and \(B\). In particular, the
commutator~\eqref{eq:heisenberg-comm-basic} is also well-defined, but
what is about its importance?

It is easy to argue that noncommutativity of observables is not an
essential prerequisite for quantum mechanics: there are constructions
of quantum theory which do not relay on it. The most prominent example
is the Feynman path integral. To focus on the really cardinal
moments, we firstly take the popular lectures~\cite{Feynman1990qed},
which present the main elements in a very enlightening way. Feynman
managed to tell the fundamental features of quantum electrodynamics
without any reference to (non-)commutativity: the entire text does not
mention it at all.

Is this an artefact of the popular nature of these lecture? Take the
academic presentation of path integral technique given in
\cite{FeynHibbs65}. It mentioned (non-)com\-mu\-ta\-ti\-vi\-ty only on
pages~115--6, 176. In addition, page~355 contains a remark on
noncommutativity of quaternions, which is irrelevant to our topic.
Moreover, page~176 highlights that noncommutativity of quantum
observables is a consequence of the path integral formalism rather
than an indispensable axiom.

But what is the mathematical source of quantum theory if noncommutativity
is not? The vivid presentation in~\cite{Feynman1990qed} uses stopwatch
with a single hand to explain the calculation of path
integrals. The angle of stopwatch's hand presents the \emph{phase} 
for a path \(x(t)\) between two points in the configuration space.
The mathematical expression for the path's phase is
\cite{FeynHibbs65}*{(2-15)}:
\begin{equation}
  \label{eq:phase-path}
  \phi[x(t)]=\mathrm{const}\cdot e^{(\rmi/\myhbar)S[x(t)]},
\end{equation}
where \(S[x(t)]\) is the \emph{classic action} along the path
\(x(t)\). Summing up contributions~\eqref{eq:phase-path} along all
paths between two points \(a\) and \(b\) we obtain the amplitude
\(K(a,b)\). This amplitude presents very accurate description of many
quantum phenomena. Therefore, expression~\eqref{eq:phase-path} is also
a strong contestant for the r\^ole of the cornerstone of quantum
theory.

Is there anything common between two ``principal''
identities~\eqref{eq:heisenberg-comm-basic} and~\eqref{eq:phase-path}?
Seemingly, not. A more attentive reader may say that there are only two
common elements there (in order of believed significance):
\begin{enumerate}
\item The non-zero Planck constant \(\myhbar\).
\item The imaginary unit \(\rmi\).
\end{enumerate}

The Planck constant was the first manifestation of quantum (discrete) 
behaviour and it is at the heart of the whole theory. In contrast, classical
mechanics is oftenly obtained as a semiclassical limit \(\myhbar
\rightarrow 0\). Thus, the non-zero Planck constant looks like a clear
marker of quantum world in its opposition to the classical
one. Regrettably, there is a common practice to ``chose our units such
that \(\myhbar=1\)''. Thus, the Planck constant becomes oftenly invisible in many
formulae even being implicitly present there. Note also, that \(1\) in
the identity \(\myhbar=1\) is not a scalar but a physical quantity
with the dimensionality of the action. Thus, the simple omission of the Planck
constant invalidates dimensionalities of physical equations.

The complex imaginary unit is also a mandatory element of quantum
mechanics in all its possible formulations. It is enough to point out
that the popular lectures~\cite{Feynman1990qed} managed to avoid any
noncommutativity issues but did mention complex numbers both
explicitly (see the Index there) and implicitly (as rotations of the
hand of a stopwatch). However, it is a common perception that complex
numbers are a useful but manly technical tool in quantum theory.

\section{Quantum Mechanics from the Heisenberg Group}
\label{sec:heisenberg-group}

Looking for a source of quantum theory we again return to the
Heisenberg commutation relations~\eqref{eq:heisenberg-comm-basic}:
they are an important part of quantum mechanics (either as a
prerequisite or as a consequence). It was observed for a long time
that these relations are a representation of the structural identities
of the Lie algebra of the Heisenberg
group~\cites{Folland89,Howe80a,Howe80b}. In the simplest case of
one dimension, the Heisenberg group \(\Space{H}{1}\) can
be realised by the Euclidean space \(\Space{R}{3}\) with the group
law:
\begin{equation}
  \label{eq:H-n-group-law}
  \textstyle
  (s,x,y)*(s',x',y')=(s+s'+\frac{1}{2}\omega(x,y;x',y'),x+x',y+y'),
\end{equation} 
where \(\omega\) is the
\emph{symplectic form} on \(\Space{R}{2}\)~\cite{Arnold91}*{\S~37}:
\begin{equation}
  \label{eq:symplectic-form}
  \omega(x,y;x',y')=xy'-x'y.
\end{equation}
Here, like for the path integral, we see another example of a quantum notion
being defined through a classical object. 

The Heisenberg group is noncommutative since \(\omega(x,y;x',y')
=-\omega(x',y';x,y)\). The collection of points \((s,0,0)\) forms the
centre of \(\Space{H}{1}\). We are interested in the unitary
irreducible representations (UIRs) of \(\Space{H}{1}\) in
infinite-dimensional Hilbert spaces.  For such a representation
\(\uir{}{}\), action of the centre shall be multiplication by an
unimodular complex number, i.e. \(\uir{}{}(s,0,0)=e^{2\pi\rmi\myhbar
  s} I\) for some real \(\myhbar\neq 0\).

Furthermore, the celebrated Stone--von~Neumann
theorem~\cite{Folland89}*{\S~1.5} tells that all UIRs of
\(\Space{H}{1}\) with the same value of \(\myhbar\) in complex Hilbert
spaces are unitary equivalent. In particular, this implies that any
realisation of quantum mechanics, e.g. the Schr\"odinger wave
mechanics, which provides the commutation
relations~\eqref{eq:heisenberg-comm-basic} shall be unitary equivalent
to the Heisenberg matrix mechanics based on these relations.

In particular, any UIR of \(\Space{H}{1}\) is equivalent to a
subrepresentation of the following representation on
\(\FSpace{L}{2}(\Space{R}{2})\):  
\begin{equation}
  \label{eq:stone-inf}
  \textstyle
  \uir{}{\myhbar}(s,x,y): f (q,p) \mapsto 
  e^{-2\pi\rmi(\myhbar s+qx+py)}
  f \left(q-\frac{\myhbar}{2} y, p+\frac{\myhbar}{2} x\right).
\end{equation}
Here \(\Space{R}{2}\) has the physical meaning of the classical
\emph{phase space} with \(q\) representing the coordinate in the
configurational space and \(p\)---the respective momentum. The
function \(f(q,p)\) in~\eqref{eq:stone-inf} presents a state of the
physical system as an amplitude over the phase space.  Thus the
action~\eqref{eq:stone-inf} is more intuitive and has many technical
advantages~\cites{Howe80b,Zachos02a,Folland89} in comparison with the
well-known Schr\"odinger representation, to which it is unitary
equivalent, of course.

Infinitesimal generators of the one-parameter semigroups
\(\uir{}{\myhbar}(0,x,0)\) and \(\uir{}{\myhbar}(0,0,y)\)
from~\eqref{eq:stone-inf} are the operators
\(\frac{1}{2}\myhbar\partial_p-2\pi\rmi q\) and
\(-\frac{1}{2}\myhbar\partial_q-2\pi\rmi p\). For these, we can
directly verify the identity:
\begin{displaymath}
\textstyle   [-\frac{1}{2}\myhbar\partial_q-2\pi\rmi p, 
\frac{1}{2}\myhbar\partial_p-2\pi\rmi q]= \rmi \myh,\quad
  \text{ where } \myh =2\pi\myhbar.
\end{displaymath}
Since we have a representation
of~\eqref{eq:heisenberg-comm-basic}, these operators can be used as
a model of the quantum coordinate and momentum.

For a Hamiltonian \(H(q,p)\) we can integrate the representation
\(\uir{}{\myhbar}\) with the Fourier transform \(\hat{H}(x,y)\) of
\(H(q,p)\):
\begin{displaymath}
  \tilde{H}=\int_{\Space{R}{2}} \hat{H}(x,y)\, \uir{}{\myhbar}(0,x,y)\,dx\,dy
\end{displaymath}
and obtain (possibly unbounded) operator \(\tilde{H}\) on
\(\FSpace{L}{2}(\Space{R}{2})\).  This assignment of the operator
\(\tilde{H}\) (quantum observable) to a function \(H(q,p)\) (classical
observable) is known as the Weyl quantization or a Weyl
calculus~\cite{Folland89}*{\S~2.1}.  The Hamiltonian \(\tilde{H}\)
defines the dynamics of a quantum observable \(\tilde{k}\) by the
\emph{Heisenberg equation}:
\begin{equation}
  \label{eq:Heisenberg-dynamics}
   \rmi\myh \frac{d\tilde{k}}{dt}=\tilde{H} \tilde{k} - \tilde{k} \tilde{H}.
\end{equation}
This is the well-known construction of quantum mechanics from
infinite-di\-men\-sional UIRs of the Heisenberg group, which can be
found in numerous sources~\cites{Kisil02e,Folland89,Howe80b}. 
%

\section{Classical Noncommutativity}
\label{sec:class-noncomm}

Now we are going to show that the balance of importance in quantum
theory shall be shifted from the Planck constant towards the imaginary
unit.  Namely, we describe a model of classical mechanics with a
non-zero Planck constant but with a different hypercomplex unit.
Instead of the imaginary unit with the property \(\rmi^2=-1\) we will
use the nilpotent unit \(\rmp\) such that \(\rmp^2=0\). The \emph{dual
  numbers} generated by nilpotent unit were already known for there
connections with Galilean relativity~\cites{Yaglom79,Gromov90a}---the
fundamental symmetry of classical mechanics---thus its appearance in
our discussion shall not be very surprising after all. Rather, we may
be curious why the following construction was unnoticed for such a
long time.

Another important feature of our scheme is that the classical
mechanics is presented by a noncommutative model. Therefore, it will
be a refutation of Dirac's claim about the exclusive r\^ole of
noncommutativity for quantum theory. Moreover, the model is developed
from the same Heisenberg group, which were used above to describe the
quantum mechanics.

Consider a four-dimensional algebra \(\algebra{C}\) spanned by
\(1\), \(\rmi\), \(\rmp\) and \(\rmi\rmp\).  We can define the following
representation \(\uir{}{\rmp\myh}\) of the Heisenberg group in a space
of \(\algebra{C}\)-valued smooth functions~\cites{Kisil10a,Kisil11c}:
\begin{eqnarray}
  \label{eq:dual-repres}
  \lefteqn{\uir{}{\rmp\myh}(s,x,y):\  f(q,p) \mapsto}\\
  && e^{-2\pi\rmi(xq+yp)}\left(f(q,p)
    +\rmp\myh \left(s f(q,p)
      +\frac{y}{4\pi\rmi}f'_q(q,p)-\frac{x}{4\pi\rmi}f'_p(q,p)\right)\right).
  \nonumber 
\end{eqnarray}
A simple calculation shows the representation property \(
\uir{}{\rmp\myh}(s,x,y)
\uir{}{\rmp\myh}(s',x',y')=\uir{}{\rmp\myh}((s,x,y)*(s',x',y'))\) for
the multiplication~\eqref{eq:H-n-group-law} on \(\Space{H}{1}\).
Since this is not a unitary representation in a complex-valued Hilbert
space its existence does not contradict the Stone--von~Neumann
theorem.  Both representations~\eqref{eq:stone-inf}
and~\eqref{eq:dual-repres} are \emph{noncommutative} and act on the phase
space. The important distinction is:
\begin{itemize}
\item The representation~\eqref{eq:stone-inf} is induced (in the sense
  of Mackey~\cite{Kirillov76}*{\S~13.4}) by the \emph{complex-valued} unitary character
  \(\uir{}{\myhbar}(s,0,0)=e^{2\pi\rmi\myhbar s}\) of the centre of
  \(\Space{H}{1}\).
\item The representation~\eqref{eq:dual-repres} is similarly induced
  by the \emph{dual number-valued} character
  \(\uir{}{\rmp\myh}(s,0,0)=e^{\rmp\myh s}=1+\rmp\myh s\) of the
  centre of \(\Space{H}{1}\), cf.~\cite{Kisil09c}. Here dual numbers
  are the associative and commutative two-dimensional algebra spanned
  by \(1\) and \(\rmp\).
\end{itemize}

Similarity between~\eqref{eq:stone-inf}
and~\eqref{eq:dual-repres} is even more explicit 
if~\eqref{eq:dual-repres} is written as:
\begin{equation}
  \label{eq:dual-as-SB}
  \uir{}{\myhbar}(s,x,y): f (q,p) \mapsto 
  e^{-2\pi(\rmp \myhbar s+\rmi(qx+py))}
  f \left(q-\frac{\rmi\myhbar}{2} \rmp y, p+\frac{\rmi\myhbar}{2} \rmp
    x\right).
\end{equation}
Here, for a differentiable function \(k\) of a real variable \(t\),
the expression \(k(t+\rmp a)\) is understood as \(k(t)+\rmp a k'(t)\),
where \(a\in\Space{C}{}\) is a constant. For a real-analytic function
\(k\) this can be justified through its Taylor's expansion,
see~\citelist{\cite{CatoniCannataNichelatti04}
  \cite{Zejliger34}*{\S~I.2(10)} \cite{Gromov90a}}. The same expression appears
within the non-standard analysis based on the idempotent unit
\(\rmp\)~\cite{Bell08a}.

The infinitesimal generators of one-parameter subgroups
\(\uir{}{\rmp\myh}(0,x,0)\) and \(\uir{}{\rmp\myh}(0,0,y)\) in~\eqref{eq:dual-repres} are
\begin{displaymath}
 d\uir{X}{\rmp\myh}= -2\pi\rmi q-\frac{\rmp\myh}{4\pi\rmi}\partial_p \quad \text{ and }
  \quad
 d\uir{Y}{\rmp\myh}= -2\pi\rmi p+\frac{\rmp\myh}{4\pi\rmi}\partial_q,
\end{displaymath}
respectively. We calculate their commutator:
\begin{equation}
  \label{eq:dual-classical-commutator}
  d\uir{X}{\rmp\myh}\cdot  d\uir{Y}{\rmp\myh}-
   d\uir{Y}{\rmp\myh}\cdot  d\uir{X}{\rmp\myh}=\rmp\myh.
\end{equation}
It is similar to the Heisenberg
relation~\eqref{eq:heisenberg-comm-basic}: the commutator is non-zero
and is proportional to the Planck constant. The only difference is the
replacement of the imaginary unit by the nilpotent one. The radical
nature of this change becomes clear if we integrate this representation
with the Fourier transform \(\hat{H}(x,y)\) of a Hamiltonian function
\(H(q,p)\):
\begin{equation}
  \label{eq:classical-int-represe}
  \mathring{H} =  \int_{\Space{R}{2n}}\hat{H}(x,y)\,
  \uir{}{\rmp\myh}(0,x,y)\,dx\,dy
  =H+\frac{\rmp\myh}{2} \left(\frac{\partial  H}{\partial p}\frac{\partial\  }{\partial q}
    - \frac{\partial  H}{\partial q} \frac{\partial\ }{\partial p}\right). 
\end{equation}
This is a first order differential operator on the phase space. It 
generates a dynamics of a classical observable \(k\)---a smooth real-valued
function on the phase space---through the equation isomorphic to the
Heisenberg equation~\eqref{eq:Heisenberg-dynamics}:
\begin{displaymath}
  \rmp \myh \frac{d \mathring{k}}{d t}= \mathring{H} \mathring{k} - \mathring{k} \mathring{H}.
\end{displaymath}
Making a substitution from~\eqref{eq:classical-int-represe} and using the
identity \(\rmp^2=0\) we obtain:
\begin{equation}
  \label{eq:hamilton-poisson}
  \frac{d {k}}{d t} =\frac{\partial  H}{\partial p}\frac{\partial k }{\partial q}
    - \frac{\partial  H}{\partial q} \frac{\partial k }{\partial p}.
\end{equation}
This is, of course, the \emph{Hamilton equation} of classical
mechanics based on the \emph{Poisson bracket}.  Dirac suggested, see
the paper's epigraph, that  the commutator  \emph{corresponds} to the
Poisson bracket. However, the commutator in the
representation~\eqref{eq:dual-repres} \emph{exactly is} the Poisson
bracket. 

Note also, that both the Planck constant and the nilpotent unit
disappeared from~\eqref{eq:hamilton-poisson}, however we did use the
fact \(\myh\neq 0\) to make this cancellation.  Also, the shy
disappearance of the nilpotent unit \(\rmp\) at the very last minute
can explain why its r\^ole remain unnoticed for a long time.

\section{Discussion}
\label{sec:conclusions}

This paper revises mathematical foundations of quantum and classical
mechanics and the r\^ole of hypercomplex units \(\rmi^2=-1\) and
\(\rmp^2=0\) there.  To make the consideration complete, one may wish
to consider the third logical possibility of the hyperbolic unit
\(\rmh\) with the property
\(\rmh^2=1\)~\cites{Hudson66a,Khrennikov09book,Kisil10a,Ulrych10a,Pilipchuk10a,Kisil12a,Kisil09c},
however, this is beyond the scope of the present paper.

The above discussion
provides the following observations:
\begin{enumerate}
\item Noncommutativity is not a crucial prerequisite for
  quantum theory, it can be obtained as a consequence of other
  fundamental assumptions.
\item Noncommutativity is not a distinguished feature of quantum
  theory, there are noncommutative formulations of classical mechanics
  as well.
\item The non-zero Planck constant is compatible with classical
  mechanics. Thus, there is no a necessity to consider the
  semiclassical limit \(\myhbar \rightarrow 0\), where the
  \emph{constant} has to tend to zero.
\item There is no a necessity to request that physical observables
  form an algebra, which is a physical non-sense since we cannot add two
  observables of different dimensionalities. Quantization can be
  performed by the Weyl recipe, which requires only a structure of a
  linear space in the collection of all observables with the same
  physical dimensionality.   
\item It is the imaginary unit in~\eqref{eq:heisenberg-comm-basic},
  which is ultimately responsible for most of quantum effects.
  Classical mechanics can be obtained from the similar commutator
  relation~\eqref{eq:dual-classical-commutator} using the nilpotent
  unit \(\rmp^2=0\).
\end{enumerate}
In Dirac's opinion, quantum noncommutativity was so important because
it guaranties a non-trivial commutator, which is required to
substitute the Poisson bracket. In our model, multiplication of
classical observables is also non-commutative and the Poisson bracket
exactly is the commutator. Thus, these elements do not separate quantum
and classical models anymore.

Still, Dirac may be right that we need to change a single assumption
to get a transition between classical mechanics and quantum. But, it
shall not be a move from commutative to noncommutative. Instead, we
need to replace a representation of the Heisenberg group induced from
dual number-valued character by the representation induced by the
complex-valued character. Our resume can be stated like the title of
the paper:
\begin{quote}
  Classical\(/\)Quantum\(=\)Dual numbers\(/\)Complex numbers.
\end{quote}

\section*{Acknowledgements}
I am grateful to the anonymous referee of \emph{Mathematical
  Intelligencer} for many critical remarks, which helped to improve
presentation and forced to increase the number of quotes in the paper.
Even more importantly, the referee report confirmed a necessity of
this paper to be written. 

I am also grateful to the referee of of the
journal \emph{Izvestiya Komi NC}, who suggested the
expression~\eqref{eq:dual-as-SB}. At last but not least, I express my
gratitude to N.A.~Gromov for numerous discussions of various topics
related to the idempotent unit. 

{\small
\bibliography{abbrevmr,akisil,analyse,algebra,arare,aclifford,aphysics,ageometry}
}

\newpage

\setcounter{section}{0}
\setcounter{equation}{0}
\setcounter{page}{1}
\thispagestyle{plain}
\vspace*{30mm}
  \selectlanguage{russian}
\markboth{В. В. Кисиль}{Коммутация наблюдаемых и механика}

\begin{center}
{\bfseries
 {\Large Является ли коммутация наблюдаемых главным отличием классической
  механики от квантовой?}\\[5mm]
 {\large \href{http://www.maths.leeds.ac.uk/~kisilv/}{В. В. Кисиль}}\\[3mm]}



\end{center}

\date{\today}

\begin{abstract}
  В 1926 году Дирак предположил, что квантовая механика может быть
  получена из классической заменой единственного допущения. По его
  мнению, классическая механика определяется коммутативными величинами
  (<<c-числами>>) в то время как квантовая требует некоммутативных
  (<<q-чисел>>).  Остальные допущения являются общими для обоих
  теорий. В данной работе мы критически пересматриваем предложение
  Дирака.

  С этой целью мы представляем некоммутативную модель
  \emph{классической} механики с ненулевой постоянной Планка. Это
  возможно благодаря использованию нильпотентной единицы \(\rmp\)
  такой, что \(\rmp^2=0\). Следовательно, решающую роль в построении
  квантовой теории выполняет мнимая комплексная единица.




\end{abstract}
  \selectlanguage{russian}

\vspace{1cm}
\par
\medskip
\hfill\parbox{0.6\textwidth}{ 
\selectlanguage{russian}
\footnotesize
  \ldots было воскресенье и ко мне впервые пришла мысль, что \(ab-
  ba\) может соответствовать скобке Пуассона.
 \par \hfill П.А.М. Дирак, \url{http://www.aip.org/history/ohilist/4575_1.html}
 }\par\medskip
  \selectlanguage{russian}

\section{Введение}
\label{sec:introduction}

Сейчас наблюдается возрождения интереса к основаниям квантовой теории,
которое поддержано заметным финансированием в различных странах. Одна
из причин этого интереса связана с прикладными инженерными вопросами
возникающими при работе с нано-объектами. Вычурная 
Копенгагенская
интерпретация была удовлетворительной для сравнительно небольшого
числа теоретических физиков (и может быть даже льстила их элитарному
духу). Однако, для массового освоения практикующими инженерами
хотелось бы иметь более реалистичную картину происходящего в
микромире. В поисках таких объяснений необходимо вернуться к самым
истокам квантовой теории.

В 1926 году Дирак предположил, что квантовая механика может быть
получена из классической заменой единственного допущения, см.~\cite{rDirac26a}:
\begin{quote}
  \selectlanguage{english} 
  \ldots there is one basic assumption of the
  classical theory which is false, and that if this assumption were
  removed and replaced by something more general, the whole of atomic
  theory would follow quite naturally. Until quite recently, however,
  one has had no idea of what this assumption could be.%
  \selectlanguage{russian}%
  \footnote{ <<\ldots существует одно базовое допущение в классической
    теории, которое неверно, и если это допущение удалить или заменить
    чем-то более общим, вся теория атома получилась бы естественно.
    Однако, до недавнего времени никто не подозревал, какое это может
    быть допущение.>>}
\end{quote}

Дирак предположил, что необходимое условие заключено в коммутационных
соотношениях Гейзенберга для наблюдаемых координаты и импульса
частицы~\cite{rDirac26a}*{(1)}: 
\begin{equation}
  \label{eq:heisenberg-comm-basic}
  q_r p_r-p_r q_r=\rmi \myh.
\end{equation}
Алгебраически это соотношение фиксирует некоммутативность величин
\(q_r\) и \(p_r\). Поэтому Дирак предложил~\cite{rDirac26a} гипотезу о
том, что классическая механика определяется коммутативными величинами
(<<c-числами>>, как он их назвал) в то время как квантовая требует
некоммутативных (<<q-чисел>>). Остальная часть теории, не
противоречащая предыдущему допущению, не требует изменений. Это в явном
виде подтверждено в следующей статье Дирака~\cite{rDirac26b}:
\begin{quote}
  \selectlanguage{english} 
  The new mechanics of the atom introduced by Heisenberg may be based
  on the assumption that the variables that describe a dynamical
  system do not obey the commutative law of multiplication, but
  satisfy instead certain quantum conditions.%
  \selectlanguage{russian}%
  \footnote{<<Новая механика атома, предложенная Гейзенбергом, может
    быть основана на допущении, что переменные описывающие динамику
    системы не следуют закону коммутативности умножения, вместо этого
    удовлетворяют некоторым квантовым соотношениям.>>}
\end{quote}
Эта же точка зрения неоднократно выражалась и в более поздних работах
\citelist{\cite{rDiracDirections}*{p.~6 (стр.~11 русского перевода)}
\cite{rDiracPrinciplesQM}*{p.~26 (стр.~41 русского перевода); }}.

Точка зрения Дирака получила широкое распространение, особенно среди
математически ориентированных учёных. Более того, расплывчатая
вариация <<квантовое---это что-то такое некоммутативное>>  изначального
предложения было с лёгкостью обращено во <<всякое некоммутативное---это
квантовое>>. Например, стало модным называть всякую некоммутативную
алгебру <<квантовым пространством>> (``quantum
space'')~\cite{rCuntz01a}. 

Давайте внимательно разберём, действительно ли некоммутативность
является важнейшим источником квантовой теории.

\section{<<Алгебра>> наблюдаемых}
\label{sec:algebra-observables}

Отбросив предположение о коммутативности наблюдаемых Дирак делает
следующее, казалось бы очень гибкое, допущение~\cite{rDirac26a}:
\begin{quote}
  \selectlanguage{english} 
  All one knows about q-numbers is that if \(z_1\) and \(z_2\) are two
  q-numbers, or one q-number and one c-number, there exist the numbers
  \(z_1 + z_2\), \(z_1 z_2\), \(z_2 z_1\), which will in general be
  q-numbers but may be c-numbers.%
  \selectlanguage{russian}%
  \footnote{<<Всё, что мы знаем о q-числах это,  если \(z_1\) и
    \(z_2\)---два 
  q-числа, или одно q-число и одно c-число, тогда существуют величины
  \(z_1 + z_2\), \(z_1 z_2\), \(z_2 z_1\), которые в общем случае являются
  q-числами, но могут оказаться и c-числами.>>}
\end{quote}
Математически это предположение (совместно с некоторыми естественными
соотношениями) обозначает, что наблюдаемые образуют алгебраическую структуру
называемую \emph{кольцом}. Далее, линейный \emph{принцип суперпозиции}
требует, что бы у наблюдаемых была так же структура векторного
пространства, что вместе с предыдущем условием характеризует множество
всех наблюдаемых как \emph{алгебру}. Некоторые работы, ориентированные
в первую очередь на математиков,
см.~\cite{rFaddeevYakubovskii09}*{\S~1.2}, прямо говорят об <<алгебре
наблюдаемых>>, что мало отличается от предыдущей цитаты
из~\cite{rDirac26a}. Это так же следует из двух взаимосвязанных допущений,
содержащихся в каноническом учебнике Дирака, на котором выросло не
одно поколение исследователей:
\begin{enumerate}
\item <<линейные операторы соответствуют динамическим
  переменным>>~\cite{rDiracPrinciplesQM}*{\S~7, p.~26, стр.~40 русского
  перевода}. 
\item <<линейные операторы можно
  складывать>>~\cite{rDiracPrinciplesQM}*{\S~7, p.~23, стр. 38 русского
  перевода}. 
\end{enumerate}

Однако, предположение, что любые две наблюдаемые допускают сложение,
полностью несовместимо с их физическим смыслом. Что бы сложение было
возможно обе наблюдаемые должны иметь одну размерность. Это тщательно
объясняют ученикам средних школ (<<нельзя складывать сапоги с
яблоками>>, как говорил мой учитель физики), но зачастую требуют
забыть в ВУЗовских учебниках.  Поэтому, стоит немного задержаться на
этом элементарном вопросе. Например, для наблюдаемых координаты \(q\)
и импульса \(p\), какова должна быть размерность выражения \(q+p\)?
Метры или \(\frac{\text{кг}\times\text{м}}{\text{сек}}\)? Если наши
расчёты показывают значение \(5\) для \(p+q\) в метрической системе,
каков будет результат при переходе к аршинам и пудам? Так как такие
вопросы не допускают внятного ответа, то предположение Дирака не
совместимо с физическим смыслом теории.

Другое распространённое определение, хромающее на ту же ногу, часто
используется в хороших книгах написанных отличными математиками,
см. например \citelist{\cite{rMackey63}*{\S~2-2}
  \cite{rFolland89}*{\S~1.1}}. Оно вводит квантовые наблюдаемые как
проекторно-значные меры на \emph{безразмерной} действительной прямой.
Такое определение немедленно влечёт (посредством функционального
исчисления операторов) существование новых наблюдаемых заданных
алгебраическими выражениями~\cite{rMackey63}*{\S~2-2, p.~63}:
\begin{quote}
  \selectlanguage{english} 
  Because of Axiom III, expressions such as \(A^2\), \(A^3+A\),
  \(1-A\), and \(e^A\) all make sense whenever \(A\) is an
  observable.%
  \selectlanguage{russian}%
  \footnote{<<Вследствие Аксиомы III, выражения вроде \(A^2\), \(A^3+A\),
  \(1-A\) и \(e^A\) все имеют смысл если \(A\) является наблюдаемой.>>}
\end{quote}
Однако, если \(A\) не является безразмерной величиной то выражение
\(A^3+A\) не может иметь никакую согласованную с этим размерность.

Конечно, физические дефекты этих (безупречных в математическом
отношении) построений не мешают физикам получать правильные ответы,
которые прекрасно согласуются с экспериментом. Нет смысла обсуждать
какими способами это достигается. Более полезно попытаться обозначить
математические основания, которые не будут страдать описанными
недостатками.

\section{Несущественная некоммутативность}
\label{sec:essent-noncomm}

Хотя мы можем складывать только наблюдаемые одной и той же
размерности, нет никаких ограничений такого рода на умножение
физических величин. Естественно, размерность произведения равна
произведению размерностей сомножителей, поэтому коммутатор
\([A,B]=AB-BA\)  всегда определён для произвольных величин
\(A\) и \(B\). В частности, коммутатор~\eqref{eq:heisenberg-comm-basic}
вполне определён. Но так ли он важен для построения квантовой
механики? 

Можно утверждать, что некоммутативность физических величин не является
\emph{необходимым предпосылкой} для оснований квантовой теории:
хорошо известны схемы обходящиеся без этого. Наиболее выдающийся
пример---интеграл по путям развитый Фейнманом (и предложенный, опять
же, Дираком). Что бы выявить действительно существенные элементы
обратимся в начале к популярным лекциям~\cite{rFeynman1990qed}, которые
представляют основу метода в очень доступной форме. Фейнман смог
рассказать главные моменты квантовой электродинамики не упомянув
некоммутативность ни разу.

Может быть это просто следствие поверхностности изложения? Возьмём
вполне академический учебник \cite{rFeynHibbs65}. В нём
некоммутативность упоминается лишь на страницах~115--6 (\S~5-3) и 176
(\S~7-3). В дополнение, на странице~355 (\S~12-10) упоминается
некоммутативность кватернионов, но это не относится к нашему
обсуждению. Более того, на странице~176 подчёркивается, что
некоммутативность квантовых величин является \emph{следствием} техники
интегрирования по путям, а не самоценной аксиомой.

Что же является математическим основанием квантовой теории, если
некоммутативность не так важна? Наглядное повествование
в~\cite{rFeynman1990qed} использует секундомер для исчисления квантовой
амплитуды. Угол поворота стрелки секундомера представляет  \emph{фазу}
для пути \(x(t)\) между двумя точками конфигурационного
пространства. Математическое выражение для этой фазы мы можем найти в 
\cite{rFeynHibbs65}*{(2-15)}:
\begin{equation}
  \label{eq:phase-path}
  \phi[x(t)]=\mathrm{const}\cdot e^{(\rmi/\myhbar)S[x(t)]},
\end{equation}
где \(S[x(t)]\)---\emph{классическое действие} вдоль пути \(x(t)\).
Сложив все вклады вида~\eqref{eq:phase-path} вдоль всех возможных
путей\footnote{Мы здесь не касаемся вопроса, каким образом можно
  математически безупречно обосновать эту процедуру.} между двумя
точками \(a\) и \(b\) мы получаем амплитуду перехода \(K(a,b)\). Эта
величина содержит в себе очень аккуратное описание многих квантовых
эффектов. Поэтому выражение~\eqref{eq:phase-path} также претендует на
роль краеугольного камня квантовой теории.

Но есть ли хоть что-то общее между двумя \emph{основополагающими}
формулами~\eqref{eq:heisenberg-comm-basic} и~\eqref{eq:phase-path}? На
первый взгляд, нет. При более детальном рассмотрении можно заметить,
что есть только два общих элемента. Перечисленные в порядке значимости
(каковой она зачастую представляется) это:
\begin{enumerate}
\item Ненулевая постоянная Планка \(\myhbar\).
\item Мнимая единица \(\rmi\).
\end{enumerate}

Действительно, постоянная Планка была исторически первой
характеристикой квантового (дискретного) поведения и вне всякого
сомнения принадлежит ядру всей теории. Более того, классическая
механика зачастую мыслится как переход от точной квантовой теории в
полуклассическом пределе \(\myhbar \rightarrow 0\). Поэтому, ненулевая
постоянная Планка считается явным признаком квантового мира в его
оппозиции к классической механике.  К сожалению, широко распространена
традиция <<выбирать такую систему единиц, в которой
\(\myhbar=1\)>>. В результате, постоянная Планка исчезает из
многих формул, где её присутствие было важно.  Отметим так же, что \(1\)
в равенстве \(\myhbar=1\) не является безразмерным скаляром, но
физической величиной с размерностью действия. Следовательно, простая
экономия на опускании этой постоянной нарушает размерность всех
физических тождеств.

Мнимая единица так же является непременным участником любых
формулировок квантовой теории. Достаточно указать, что популярные
лекции~\cite{rFeynman1990qed} прекрасно обходятся без всякого
упоминания некоммутативности, но содержат комплексные числа как явно
(см. указатель в этой книге), так и неявно---вращение стрелки
секундомера наглядно изображает изменение унимодулярной комплексной
фазы~\eqref{eq:phase-path} вдоль пути. И это второе (неявное, но очень
существенное) использование комплексных чисел даже важнее их краткого
явного упоминания. Тем не менее, комплексные числа зачастую
воспринимаются как полезный, но всё же \emph{чисто технический} элемент
теории. 

\section{Квантовая механика и группа Гейзенберга}
\label{sec:heisenberg-group}

В поисках источника квантовой теории мы вновь возвращаемся к
коммутационным соотношениям~\eqref{eq:heisenberg-comm-basic}: или в
роли необходимой аксиомы, или как важное следствие, но они являются
обязательным элементом теории. Достаточно давно стало понятно, что
эти соотношения являются представлением структурных тождеств для алгебры
Ли группы Гейзенберга~\cites{rFolland89,rHowe80a,rHowe80b}. В простейшем
случае одного измерения, группа Гейзенберга \(\Space{H}{1}\)
представляется Евклидовым пространством \(\Space{R}{3}\) с такой
групповой операцией:
\begin{equation}
  \label{eq:H-n-group-law}
  \textstyle
  (s,x,y)*(s',x',y')=(s+s'+\frac{1}{2}\omega(x,y;x',y'),x+x',y+y'),
\end{equation} 
где \(\omega\) является
\emph{симплектической формой} на \(\Space{R}{2}\)~\cite{rArnold91}*{\S~37}:
\begin{equation}
  \label{eq:symplectic-form}
  \omega(x,y;x',y')=xy'-x'y.
\end{equation}
Здесь, как и с интегралами по траекториям, мы видим ещё один пример
квантового объекта определённого в терминах классического.

Группа Гейзенберга некоммутативна в следствии косо-симметричности
симплектической формы: \(\omega(x,y;x',y')
=-\omega(x',y';x,y)\). Множество точек вида \((s,0,0)\) образует центр
\(\Space{H}{1}\). Нам потребуются унитарные неприводимы представления
\(\Space{H}{1}\) в бесконечномерных пространствах.  В таком представлении
\(\uir{}{}\) центр группы должен действовать умножением на комплексное
число с единичным модулем, то есть \(\uir{}{}(s,0,0)=e^{2\pi\rmi\myhbar
  s} I\) для некоторого \(\myhbar\neq 0\).

Далее, важная теорема Стоуна---фон  Неймана~\cite{rFolland89}*{\S~1.5}
устанавливает, что все унитарные неприводимые представления группы
\(\Space{H}{1}\) с общим значением \(\myhbar\) унитарно
эквивалентны. Из этого следует, что любые реализации
квантовой механики представляющие
соотношение~\eqref{eq:heisenberg-comm-basic} (к примеру, волновая
механика Шредингера) унитарно эквивалентна  матричной механике
Гейзенберга. 

В частности, любое унитарное неприводимое представление
\(\Space{H}{1}\) эквивалентно подпредставлению следующего
представления в пространстве \(\FSpace{L}{2}(\Space{R}{2})\):
\begin{equation}
  \label{eq:stone-inf}
  \textstyle
  \uir{}{\myhbar}(s,x,y): f (q,p) \mapsto 
  e^{-2\pi\rmi(\myhbar s+qx+py)}
  f \left(q-\frac{\myhbar}{2} y, p+\frac{\myhbar}{2} x\right).
\end{equation}
Здесь \(\Space{R}{2}\) может быть отождествленно с классическим
\emph{фазовым пространством}, где \(q\) обозначает координату в
конфигурационном пространстве и \(p\)---со\-от\-вет\-ству\-ю\-щий
импульс.  Функция \(f(q,p)\) в~\eqref{eq:stone-inf} представляет
состояние физической системы как амплитуду на фазовом пространстве.
Поэтому, по сравнению с более известным представлением Шрёдингера на
действительной оси (конфигурационном пространстве),
представление~\eqref{eq:stone-inf} более интуитивно и имеет много
технических достоинств~\cites{rHowe80b,rZachos02a,rFolland89}. Хотя, как
было отмечено выше, оба представление унитарно эквивалентны.

Инфинитезимальные порождающие одно-параметрических подгрупп
\(\uir{}{\myhbar}(0,x,0)\) и \(\uir{}{\myhbar}(0,0,y)\)
в~\eqref{eq:stone-inf} есть операторы
\(\frac{1}{2}\myhbar\partial_p-2\pi\rmi q\) и
\(-\frac{1}{2}\myhbar\partial_q-2\pi\rmi p\). Для них непосредственно
проверяется тождество:
\begin{displaymath}
\textstyle   [-\frac{1}{2}\myhbar\partial_q-2\pi\rmi p, 
\frac{1}{2}\myhbar\partial_p-2\pi\rmi q]= \rmi \myh,\quad
  \text{ где } \myh =2\pi\myhbar.
\end{displaymath}
Так как мы имеем представление
тождества~\eqref{eq:heisenberg-comm-basic}, эти операторы могут
использоваться как представители квантовых наблюдаемых координаты и
импульса.

Имея  классический гамильтониан \(H(q,p)\), мы можем проинтегрировать
его преобразование Фурье \(\hat{H}(x,y)\) с представлением
\(\uir{}{\myhbar}\) :
\begin{displaymath}
  \tilde{H}=\int_{\Space{R}{2}} \hat{H}(x,y)\, \uir{}{\myhbar}(0,x,y)\,dx\,dy
\end{displaymath}
и получим (возможно неограниченный) оператор \(\tilde{H}\) на
\(\FSpace{L}{2}(\Space{R}{2})\).  Такое соответствие оператора  
\(\tilde{H}\) (квантовой наблюдаемой) к функции \(H(q,p)\)
(классической наблюдаемой) известно как \emph{квантование Вейля} или
\emph{исчисление Вейля}~\cite{rFolland89}*{\S~2.1}.  Гамильтониан \(\tilde{H}\)
определяет динамику квантовой наблюдаемой \(\tilde{k}\) через
\emph{уравнение Гейзенберга}:
\begin{equation}
  \label{eq:Heisenberg-dynamics}
   \rmi\myh \frac{d\tilde{k}}{dt}=\tilde{H} \tilde{k} - \tilde{k} \tilde{H}.
\end{equation}
Такое построение квантовой механики на основе унитарных неприводимых
представлений группы Гейзенберга хорошо известно,
см. например~\cites{rKisil02e,rFolland89,rHowe80b}.  

\section{Классическая некоммутативность}
\label{sec:class-noncomm}

Сейчас мы покажем, что в квантовой теории по-настоящему важным
являются комплексные числа, а вовсе не ненулевая постоянная Планка,
как принято думать. Конкретно, мы представим модель классической
механики с ненулевой постоянной Планка, но с другими гиперкомплексными
числами. Вместо мнимой единицы \(\rmi\) со свойством \(\rmi^2=-1\) мы
используем нильпотентную единицу \(\rmp\) такую, что \(\rmp^2=0\).
Хорошо известно, что порождённые ей \emph{дуальные числа} связаны с
относительностью Галилея~\cites{rYaglom79,rGromov90a}---важной
симметрией классической механики---так что её появление в нашем
исследовании не такая уж и неожиданность. Скорее, мы должны
удивляться, почему дуальные числа так мало известны и так редко
используются в современной физике (да и математике).

Другой важной особенностью нашей модели классической механики является
её некоммутативность. Таким образом, она опровергает предположение
Дирака о некоммутативности как важнейшем источнике всех квантовых
построений. Более того, наша модель будет выведена из всё той же
группы Гейзенберга, что ещё больше роднит квантовую и классическую
теории.

Рассмотрим четырёхмерную алгебру \(\algebra{C}\) с базисом
\(1\), \(\rmi\), \(\rmp\) и \(\rmi\rmp\).  Можно определить следующее
представление \(\uir{}{\rmp\myh}\) группы Гейзенберга в пространствo
\(\algebra{C}\)-значных гладких функций~\cites{rKisil10a,rKisil11c}:
\begin{eqnarray}
  \label{eq:dual-repres}
  \lefteqn{\uir{}{\rmp\myh}(s,x,y):\  f(q,p) \mapsto}\\
  && e^{-2\pi\rmi(xq+yp)}\left(f(q,p)
    +\rmp\myh \left(s f(q,p)
      +\frac{y}{4\pi\rmi}f'_q(q,p)-\frac{x}{4\pi\rmi}f'_p(q,p)\right)\right).
  \nonumber 
\end{eqnarray}
Непосредственно проверяется тождество
\begin{displaymath}
  \uir{}{\rmp\myh}(s,x,y)
\uir{}{\rmp\myh}(s',x',y')=\uir{}{\rmp\myh}((s,x,y)*(s',x',y'))
\end{displaymath}
для группового умножения~\eqref{eq:H-n-group-law} на \(\Space{H}{1}\).
Так как \(\uir{}{\rmp\myh}\) не является унитарным представлением в
комплексном векторном пространстве, то оно не подпадает под действие
теоремы Стоуна--фон Неймана.  Оба представления~\eqref{eq:stone-inf}
и~\eqref{eq:dual-repres} являются \emph{некоммутативными} и действуют
на функциях заданных на фазовом пространстве. Важное отличие между
этими двумя представлениями таково:
\begin{itemize}
\item Представление~\eqref{eq:stone-inf} индуцировано (в смысле
  Макки~\cite{rKirillov76}*{\S~13.4})  \emph{комплекснозначным} характером
  \(\uir{}{\myhbar}(s,0,0)=e^{2\pi\rmi\myhbar s}\)
  центра группы
  \(\Space{H}{1}\).
\item Представление~\eqref{eq:dual-repres} сходным образом
  индуцировано характером в \emph{дуальных числах}
  \(\uir{}{\rmp\myh}(s,0,0)=e^{\rmp\myh s}=1+\rmp\myh s\) центра
  \(\Space{H}{1}\), ср.~\cite{rKisil09c}. (Дуальные числа образуют
  двумерную коммутативную ассоциативную алгебру с базисом \(\{1,
  \rmp\}\).)
\end{itemize}

Сходство представлений~\eqref{eq:stone-inf} и~\eqref{eq:dual-repres}
будет ещё более наглядным 
если записать~\eqref{eq:dual-repres} в виде:
\begin{equation}
  \label{eq:dual-as-SB}
  \uir{}{\myhbar}(s,x,y): f (q,p) \mapsto 
  e^{-2\pi(\rmp \myhbar s+\rmi(qx+py))}
  f \left(q-\frac{\rmi\myhbar}{2} \rmp y, p+\frac{\rmi\myhbar}{2} \rmp
    x\right).
\end{equation}
Здесь, для дифференцируемой функции \(k\)
действительной переменной \(t\), выражение  \(k(t+\rmp a)\) понимается как \(k(t)+\rmp a k'(t)\) при произвольной константе \(a\in\Space{C}{}\).
Для аналитических функций действительной переменной это может быть
обосновано через их разложение в ряд
Тейлора~\citelist{\cite{rCatoniCannataNichelatti04} 
  \cite{rZejliger34}*{\S~I.2(10)} \cite{rGromov90a}}. Родственный источник этого
выражения находится также в варианте нестандартного анализа
использующего идемпотентную единицу \(\rmp\)~\cite{rBell08a}.

Инфинитезимальные порождающие для одномерных подгрупп
\(\uir{}{\rmp\myh}(0,x,0)\) и \(\uir{}{\rmp\myh}(0,0,y)\) в
представлении~\eqref{eq:dual-repres} соответственно есть: 
\begin{displaymath}
 d\uir{X}{\rmp\myh}= -2\pi\rmi q-\frac{\rmp\myh}{4\pi\rmi}\partial_p \quad \text{ и }
  \quad
 d\uir{Y}{\rmp\myh}= -2\pi\rmi p+\frac{\rmp\myh}{4\pi\rmi}\partial_q.
\end{displaymath}
Непосредственно вычисляется их коммутатор:
\begin{equation}
  \label{eq:dual-classical-commutator}
  d\uir{X}{\rmp\myh}\cdot d\uir{Y}{\rmp\myh}-
   d\uir{Y}{\rmp\myh}\cdot  d\uir{X}{\rmp\myh}=\rmp\myh.
\end{equation}
Это тождество похоже на коммутационные соотношение
Гейзенберга~\eqref{eq:heisenberg-comm-basic}:  коммутатор
отличен от нуля и пропорционален постоянной Планка. Единственное
различие заключается в замене мнимой единицы не нильпотентную. Природа
этого замещения проявится когда мы проинтегрируем
представление~\eqref{eq:dual-repres} с преобразованием Фурье
\(\hat{H}(x,y)\) гамильтониана \(H(q,p)\):
\begin{equation}
  \label{eq:classical-int-represe}
  \mathring{H} =  \int_{\Space{R}{2n}}\hat{H}(x,y)\,
  \uir{}{\rmp\myh}(0,x,y)\,dx\,dy
  =H+\frac{\rmp\myh}{2} \left(\frac{\partial  H}{\partial p}\frac{\partial\  }{\partial q}
    - \frac{\partial  H}{\partial q} \frac{\partial\ }{\partial p}\right). 
\end{equation}
Мы получили дифференциальный оператор первого порядка на фазовом
пространстве. Такой оператор порождает динамику классической
наблюдаемой  \(k\)---гладкой вещественно-значной функции на фазовом
пространстве---посредством уравнения сходного с уравнением
Гейзенберга~\eqref{eq:Heisenberg-dynamics}: 
\begin{displaymath}
  \rmp \myh \frac{d \mathring{k}}{d t}= \mathring{H} \mathring{k} - \mathring{k} \mathring{H}.
\end{displaymath}
Подставляя~\eqref{eq:classical-int-represe} и используя
тождество \(\rmp^2=0\) мы получаем: 
\begin{equation}
  \label{eq:hamilton-poisson}
  \frac{d {k}}{d t} =\frac{\partial  H}{\partial p}\frac{\partial k }{\partial q}
    - \frac{\partial  H}{\partial q} \frac{\partial k }{\partial p}.
\end{equation}
Это, естественно---\emph{уравнение Гамильтона} в классической механике
использующее \emph{скобку Пуассона} двух функций \(H\) и \(k\). Как
отмечено в эпиграфе, Дирак предположил, что коммутатор должен
\emph{соответствовать} скобке Пуассона. Однако, мы обнаружили, что
коммутатор в представлении~\eqref{eq:dual-repres} \emph{в точности
  является} скобкой Пуассона.

Отметим, что и постоянная Планка, и нильпотентная единица сокращаются из
окончательного уравнения~\eqref{eq:hamilton-poisson}, но для этого
преобразования было важно, что \(\myh\neq 0\). Такое застенчивое
исчезновение в самый последний момент может объяснить, почему
величины \(\myh\) и \(\rmp\) обычно остаются незамеченными в
классической теории.

\section{Заключение}
\label{sec:conclusions}

В данной работе мы пересматриваем математические основания квантовой и
классической механики, а так же роль гиперкомплексных единиц
\(\rmi^2=-1\) и \(\rmp^2=0\) в этих теориях. Для того, что бы сделать
рассмотрение полным мы должны упомянуть третью логическую
возможность: гиперболическую единицу \(\rmh\) со свойством
\(\rmh^2=+1\),
см.~\cites{rHudson66a,rKhrennikov09book,rKisil10a,rUlrych10a,rPilipchuk10a,rKisil12a,rKisil09c},
однако её обсуждение выходит за рамки данной статьи.

Сделанный анализ приводит к следующим заключениям:
\begin{enumerate}
\item Некоммутативность не обязательно включать в аксиоматизацию
  квантовой теории, она естественно получается как следствие других
  базовых предпосылок.
\item Некоммутативность не является отличительной чертой квантовой
  теории от классической, существуют некоммутативные модели
  классической механики.
\item Ненулевая постоянная Планка вполне совместима с классической
  механикой. Нет никакой необходимости рассматривать полуклассический
  предел \(\myhbar \rightarrow 0\), в котором \emph{константа} должна
  \emph{стремится} к нулю.
\item Нет никакой необходимости рассматривать множество наблюдаемых
  как алгебру, что несовместимо с базовыми физическим смыслом теории.
  Квантование можно производить по процедуре Вейля, которое требует от
  множества наблюдаемых с одной физической размерностью всего лишь
  структуры векторного пространства.
\item Решающую роль в построении любой квантово-механической модели
  играет комплексная мнимая единица,
  см.~\eqref{eq:heisenberg-comm-basic} и~\eqref{eq:phase-path}.
  Классическая механика может быть получена заменой мнимой единицы на
  нильпотентную \(\rmp^2=0\) в коммутационных
  соотношениях~\eqref{eq:dual-classical-commutator}.
\end{enumerate}
Заметим, что некоммутативность играла такую важную роль в построениях
Дирака, потому что нетривиальный коммутатор требовался, как замена
классической скобке Пуассона. Мы показали, что умножение классических
наблюдаемых тоже может быть некоммутативным и в этом случае коммутатор
в точности совпадает со скобкой Пуассона. Таким образом, водораздел
между двумя теориями \emph{не} проходит по линии коммутативно/некоммутативно.

Возможно, Дирак всё-таки прав предполагая, что есть всего лишь одно
допущение, которое отделяет квантовую теорию от классической,
см. его первую цитату в начале статьи. Мы можем разделить его в такой
форме:
\begin{quote}
  Квантовая механика частицы основана на использовании мнимой единицы для
  индуцирования представления группы Гейзенберга с её центра. При
  замене мнимой единицы на нильпотентную получаем классическую
  механику, причём все остальные компоненты теории (некоммутативность,
  ненулевая постоянная Планка, динамическое уравнение основанное на
  коммутаторе) остаются неизменным.
\end{quote}

\section*{Благодарности}
Автор благодарен анонимному рецензенту журнала  {\emph{Mathematical
  Intelligencer}} за критический отзыв, который позволил немного улучшить
изложение в данной статье и потребовал заметного увеличения числа
дословных цитат. Ещё более полезным отзыв оказался тем, что укрепил
уверенность автора в необходимости этой статьи. 

Я так же благодарен рецензенту журнала <<Известия Коми НЦ УрО РАН>>
который подсказал выражение~\eqref{eq:dual-as-SB} для классического
представления группы Гейзенберга. Хочу так же выразить признательность
Н.А.~Громову за полезные обсуждения различных вопросов связанных с
нильпотентной единицей.


\small
\providecommand{\noopsort}[1]{} \providecommand{\printfirst}[2]{#1}
  \providecommand{\singleletter}[1]{#1} \providecommand{\switchargs}[2]{#2#1}
  \providecommand{\irm}{\textup{I}} \providecommand{\iirm}{\textup{II}}
  \providecommand{\vrm}{\textup{V}} \providecommand{\cprime}{'}
  \providecommand{\eprint}[2]{\texttt{#2}}
  \providecommand{\myeprint}[2]{\texttt{#2}}
  \providecommand{\arXiv}[1]{\myeprint{http://arXiv.org/abs/#1}{arXiv:#1}}
  \providecommand{\doi}[1]{\href{http://dx.doi.org/#1}{doi:
  #1}}\providecommand{\CPP}{\texttt{C++}}
  \providecommand{\NoWEB}{\texttt{noweb}}
  \providecommand{\MetaPost}{\texttt{Meta}\-\texttt{Post}}
  \providecommand{\GiNaC}{\textsf{GiNaC}}
  \providecommand{\pyGiNaC}{\textsf{pyGiNaC}}
  \providecommand{\Asymptote}{\texttt{Asymptote}}

\end{document}